\documentclass[10pt]{sig-alternate-05-2015}

\usepackage{authblk}

\usepackage{graphicx}
\usepackage{subfigure}
\usepackage{epstopdf}
\usepackage{color}
\usepackage{booktabs}

\begin{document}

\toappear{Copyright is held by the author/owner(s).\\
UrbComp'16, August 14, 2016, San Francisco, USA.}

\title{Robustness and Resilience of cities around the world}

\author{Sofiane Abbar$^{\dag}$, Tahar Zanouda $^{\dag}$, Javier Borge-Holthoefer$^{\ddag}$\\
$^{\dag}$Qatar Computing Research Institute,
Hamad Bin Khalifa University\\
PO 5825, Doha, Qatar\\
$^{\ddag}$Internet Interdisciplinary Institute (IN3-UOC),
Universitat Oberta de Catalunya\\
08018 Barcelona, Catalonia, Spain\\
\{sabbar, tzanouda\}@qf.org.qa, jborgeh@uoc.edu}

\maketitle
\begin{abstract}
The concept of {\em city} or {\em urban resilience} has emerged as one of the key challenges for the next decades. As a consequence, institutions like the United Nations or Rockefeller Foundation have embraced initiatives that increase or improve it. These efforts translate into funded programs both for action ``on the ground'' and to develop quantification of resilience, under the for of an index. Ironically, on the academic side there is no clear consensus regarding how resilience should be quantified, or what it exactly refers to in the urban context. Here we attempt to link both extremes providing an example of how to exploit large, publicly available, worldwide urban datasets, to produce objective insight into one of the possible dimensions of urban resilience. We do so {\em via} well-established methods in complexity science, such as percolation theory --which has a long tradition at providing valuable information on the vulnerability in complex systems. Our findings uncover large differences among studied cities, both regarding their infrastructural fragility and the imbalances in the distribution of critical services.\end{abstract}


\section{Introduction}
Over the past years the concept of urban or city resilience has repeatedly appeared in many contexts. While there is a general agreement that it is related to the capacity of a urban area to confront uncertainty and/or risk, it is not clear how it should be operationalized or quantified. Meerow {\em et al.} \cite{meerow2016defining} identify as many as 25 different definitions of urban resilience in the literature, related more or less strictly to Engineering, Environmental Sciences, Business and Finance, or Social Sciences. Indeed, it seems common sense that ``generic resilience'' (capacity of a system to recover its initial state after a shock) can be mapped onto a myriad of particular definitions which differ only in the number of affected subsystems (infrastructural, social, ecological) and the time scale of the disturbance itself: a shock may arrive under the form of a natural disaster (fast, system-wide), an economic crisis (slow, specific) or the long-term urban footprint on the surrounding environment (very slow and hardly visible in the city itself).

\begin{figure}[tp]
 \centering
 \includegraphics[width=0.95\linewidth]{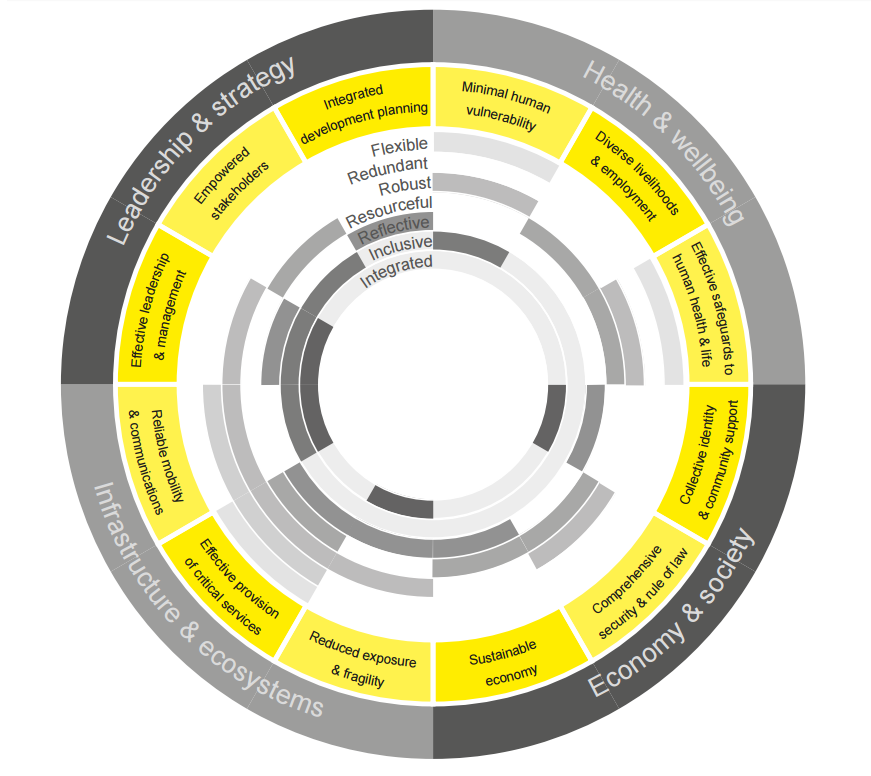}
 \caption{Guidelines to the CRI. Resilience is mapped onto 4 key dimensions (outmost ring), which are in turn subdivided into 12 specific goals.}
 \label{fig:cri}
\end{figure}

These~ semantic~ difficulties~ go~ well~ beyond~ pure academia, and can affect actual efforts towards increasing urban resilience. Public and private institutions, like the United Nations or the Rockefeller Foundation, are currently supporting initiatives which focus, explicitly or implicitly, on this topic. If, however, its definition is unclear, these programs encounter difficulties at making decisions --whether a given investment makes sense or not. In this direction, the ``100 Resilient Cities Program'' (100RC) at the Rockefeller Foundation has recently released some guidelines with regard to what makes a city resilient (City Resilience Index, CRI)\footnote{See http://www.cityresilienceindex.org/}. This effort to build an agreed {\em rationale} is interesting, because it breaks down the problem into four dimensions (see Figure \ref{fig:cri}), which is already a significant advance toward operationalization.

In this paper we focus on the infrastructural side of the 100RC, which breaks this dimension into three goals: reliable mobility and communications; reduced exposure and fragility; and effective provision of critical services. We relate each of these goals to different aspects of network robustness, which has a longstanding tradition in complex systems, and percolation theory in particular \cite{albert2000error,callaway2000network,cohen00}. Our claim is that the current availability of large digital datasets, and of solid foundations in complex networks can offer a suitable quantitative complement to the evaluation of city resilience, which --as in the example of CRI-- typically relies on general traits (e.g. exposure of a zone to flooding hazards).

Our study relates the first aforementioned goal with the robustness of the road network against random failure of its segments, inasmuch ``reliable mobility'' relates to daily connectivity between places, people and services. In this sense, the network is prone to fail anywhere, anytime (accidents, congestion, system-wide shocks), as opposed to targeted attacks on critical parts of the infrastructure which is the second goal (``reduced exposure and fragility''). Finally, we combine this network approach with large datasets on service location extracted from Foursquare \footnote{https://foursquare.com/} -- a location-based social network (LBSN) -- to understand to what extent a broken infrastructure can affect the reachability of critical services (``effective provision of critical services''), i.e. whether a city offers a balanced and redundant supply of resources. Such services include but are not limited to ``medical center'' and ``transportation infrastructure''. 
 
In order to place the paper in the right context, we have chosen to perform this work mostly on cities included in 100RC for which we could collected data, offering a study at scale which allows for a cross-city robustness comparison that has --to the best of our knowledge-- not been addressed before.

\section{Data}
\begin{figure}[tp]
 \centering
 \includegraphics[width=0.95\linewidth]{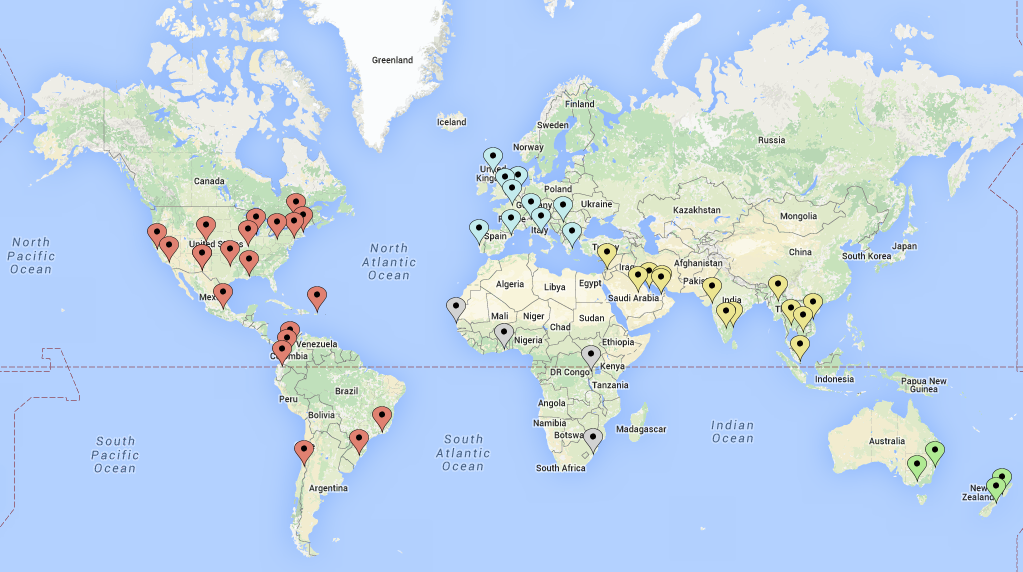}
 \caption{The geography of cities covered in this study. As we see, cities are from around the world and span six continents.}
 \label{fig:cities}
\end{figure}

To conduct our study, we focus on cities participating in the 100RC program to which we deliberately added 5 cities from the Gulf Cooperation Council (GCC) in the Middle East. These cities are Abu-Dhabi and Dubai (UAE), Doha (Qatar), Manama (Bahrain), and Riyadh (Saudi Arabia). Figure \ref{fig:cities} gives an overview of the geographical distribution of the 54 cities considered. We can easily see that the cities span all continents (except Antarctic) with 4 cities from Africa, 14 from Asia, 10 from Europe, 14 from North America, 6 from South America, 1 from Central America, and 4 from Oceania. The geographic diversity and cultural heterogeneity of these cities provide two advantages. First, it diminishes the biases one could observe in studies that focus on cities from a particular country or continent. Furthermore, it allows a more robust generalization of the findings.

For each city, we requested two types of data: the road network and the geographic distribution of different services available in the city. We describe in the following section the data used and the processes by which it was prepared. We also report some initial observations. 

\subsection{From Open Street Map to Road Networks} 
OpenStreetMap (OSM) \footnote{http://www.openstreetmap.org} is a popular open source mapping service built by a community of volunteer mappers who contribute to the tagging of cities by maintaining data about buildings, shops, roads, metro stations and any other service that has a value to be on a map. As a crowd-sourced platform, OSM relies on its active community of contributers to create accurate maps by leveraging their local knowledge on neighborhoods. Because OSM databases are exclusively contributed by volunteers, it is important to understand that its accuracy and completeness vary from a city to another. Generally speaking, large touristic cities like Paris, New York, and Singapore tend to expose data with high resolution and quality as compared to smaller and isolated cities. 

While OSM provides reasonable APIs to query its databases, we decided to use a more comprehensive online service called {\it Mapzen}\footnote{http://mapzen.com} to download ``extracts'' of relevant cities. The extracts we requested come in the form of shape files that are easy to process. {\it Mapzen} updates on a daily basis its city extracts by querying OSM. 

\subsection{Creating road networks}
The shape file of cities come with different files defining different services in the city such as transport points, building, places, administrative boundaries, and roads. All these objects are made in OSM with {\it nodes}. A node is defined with two mandatory attributes that are its {\textit ID} (unique identifier) and its \textit{GPS} coordinates (longitude, latitude), and a list of complementary information such as names, addresses, and types of locations. Complex objects such as buildings are defined as polygons of nodes whereas paths (segments of roads) are defined as lists of nodes. 

\begin{figure}[htp]
 \centering
 \includegraphics[width=0.95\linewidth]{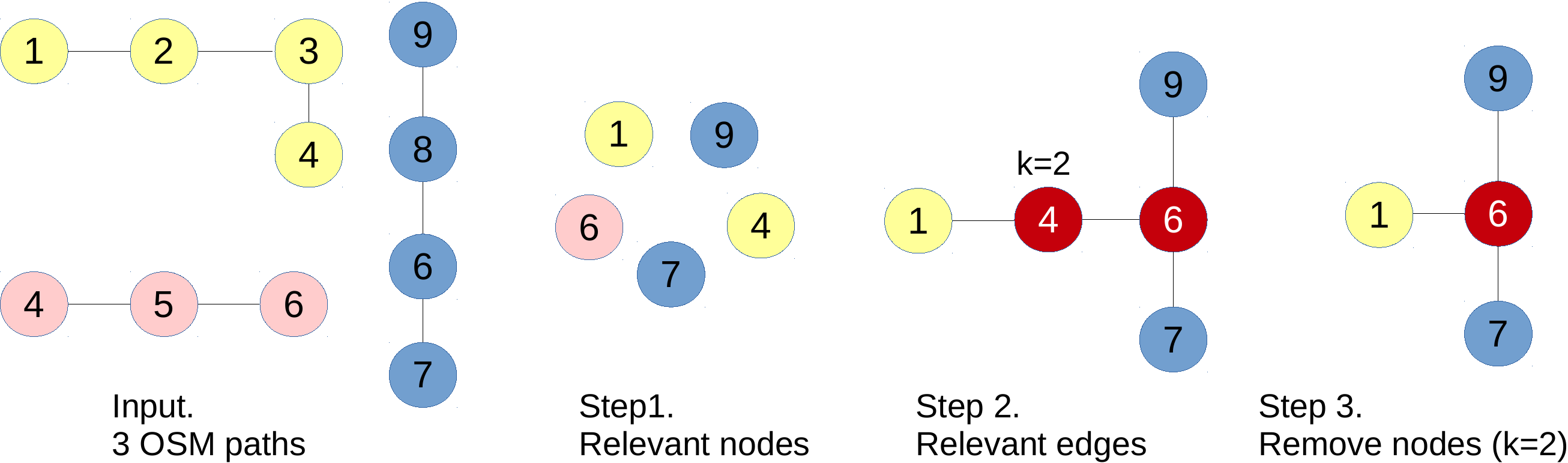}
 \caption{Building road network from OSM paths. Given three roads, we first identify the set of relevant nodes. Nodes number 1, 3, 4, 6, 7 and 9 are relevant because they start/end roads. Nodes 4 and 6 are also intersections as both belong to more than one path. But node 4 is removed in the third step at it holds a degree $k=2$. }
 \label{fig:building_road_network}
\end{figure}

Given the road shape file (i.e., list of paths) of a city, our objective is to create a real road network in which nodes are intersections and edges are road segments between them. The major challenge in this task is to know how to merge different paths (streets) into the same transportation unit (road). Different techniques have been proposed in the literature to deal with this problem. Some of them rely on the street names ({\em SN}) to group them \cite{Jiang2014SN} whereas other methods are based on the geometry of streets and look at the intersection continuity between streets ({\em ICN}) \cite{Porta2006ICN}.  
Given that both approaches have pros and cons, we decided to go with our own three-steps heuristic (see Figure \ref{fig:building_road_network} for an illustrative example.) 

{\bf Step 1.} We identify the list of relevant nodes. A node is considered relevant to the road network if and only if it verifies one of the following two conditions: (i.) The node is the beginning or the end of a path. (ii.) The node belongs to more than one path. 
The first condition captures all nodes necessary to the definition of roads whereas the second condition captures intersections between roads.

{\bf Step 2.} We iteratively scan through the list of all paths available in the city shape file. For each path, we create an edge between every successive pair of relevant nodes identified in step one. This process discards intermediate nodes that OSM mappers use primarily to shape road segments.

{\bf Step 3.} We recursively prune from the resulting graph all nodes with degree $k=2$ and replace them with an edge connecting the two nodes to which they are linked. The intuition here is that nodes with degree $k=2$ do not add any semantic to the road network as it is envisioned in this study. 

\begin{figure}
\centering
\subfigure[New York\label{fig:newyork_network}]{\includegraphics[width=.45\columnwidth]{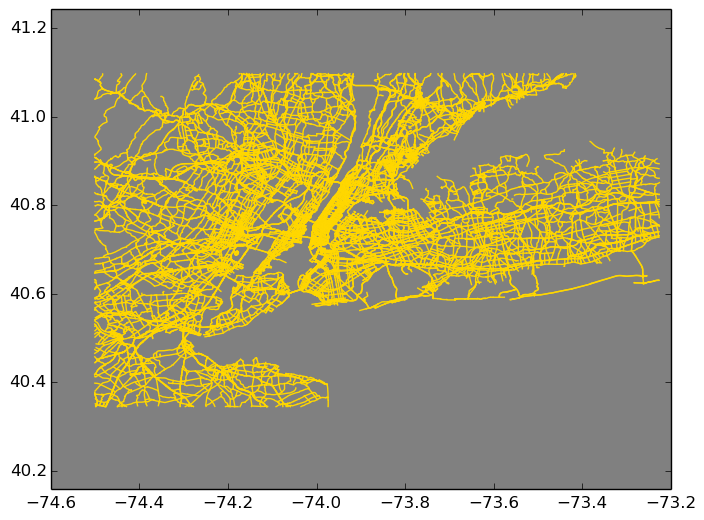}}
  \qquad
\subfigure[London\label{fig:london_network}]{\includegraphics[width=.45\columnwidth]{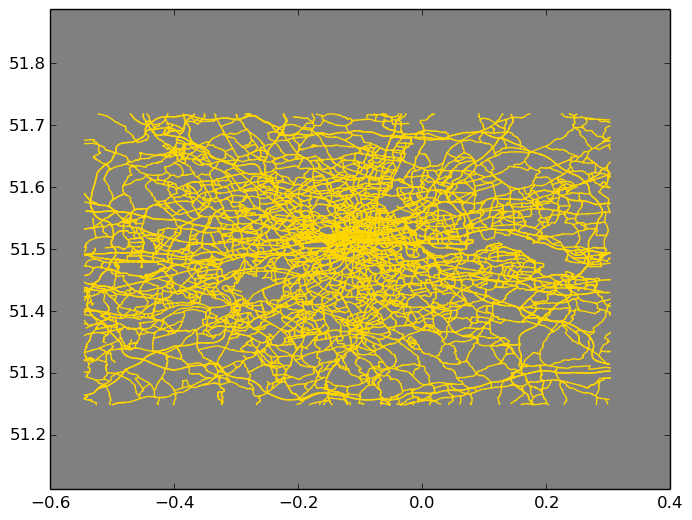}}

\subfigure[Dakar\label{fig:dakar_network}]{\includegraphics[width=.45\columnwidth]{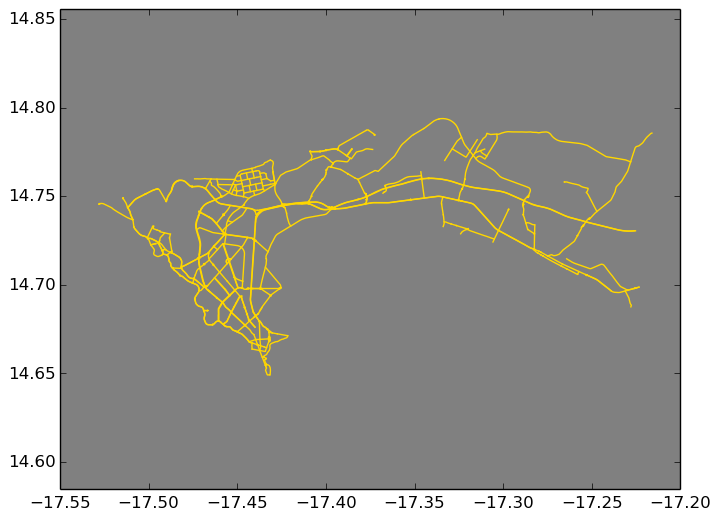}}
\qquad
\subfigure[Dubai\label{fig:dubai_network}]{\includegraphics[width=.45\columnwidth]{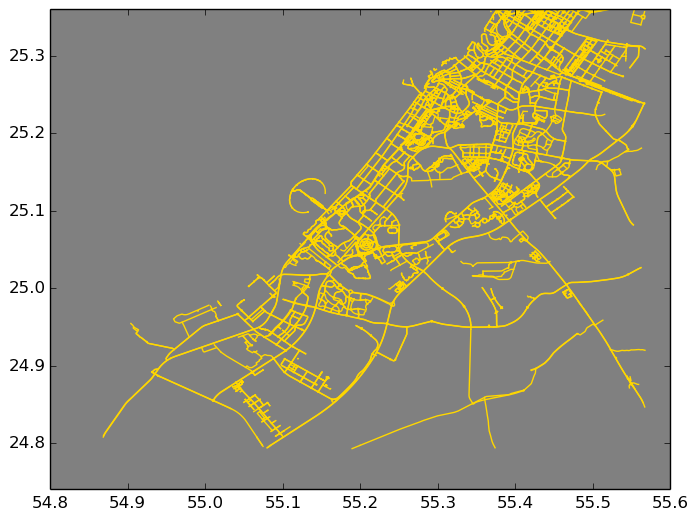}}
  
\caption{Road networks of some selected cities as extracted from OpenStreetMap. We see that cities show different types of networks in terms of size, density, and topology}
\label{fig:road_networks_eg}
\end{figure}

Examples of real road networks of four cities generated with our heuristic are shown in Figure \ref{fig:road_networks_eg}.

In Table \ref{tbl:stats} we summarize some key structural traits of the extracted road networks. For consistency reason, we reduce the road network of each city to its first giant connected component (GCC) for which we highlight $V$(city) (number of nodes), $E$(city) (number of edges), Length (total kilometers), average degree ($k$). We include also Meshness $M \in [0, 1]$, defined as the ratio of the existing facets divided by the total number of all possible facets in a graph, where $M \sim 0$ (for tree-like graphs) and $M \sim 1$ (for complete graphs) \cite{Jiaqiu2015Resilience}:
 \begin{equation}
  M = (E - V + 1) / (2V - 5)
 \end{equation}
Finally, we report the ``organic'' score to assess whether a road network has been planned or not \cite{Jiaqiu2015Resilience}, computed as the proportion of dead ends (nodes with degree $k=1$) and unfinished intersections (nodes with degree $k=3$) to the total number of nodes:
\begin{equation}
  Org = (V(k=1) + V(k=3)) / V
\end{equation}

We found that the meshness scores of cities in this study vary between $0.084$ for San-Juan and $0.228$ for Riyadh which are consistent with the scores reported in \cite{cardillo2006PlanarGraphs} on analyzing planar graphs of twenty cities. The low scores are mainly due to the absence of squares and triangles in real-world cities. Non surprisingly, organic scores are high for almost all the cities, with a noticeable advantage for new and fast growing cities such as Accra ($0.97$) and Doha ($0.96$) as compared to old cities such as Barcelona and San-Francisco ($0.89$).

\begin{table}[t]
  \caption{Statistics about road networks of different cities.}
  \label{tbl:stats}
  \includegraphics[width=0.97\linewidth]{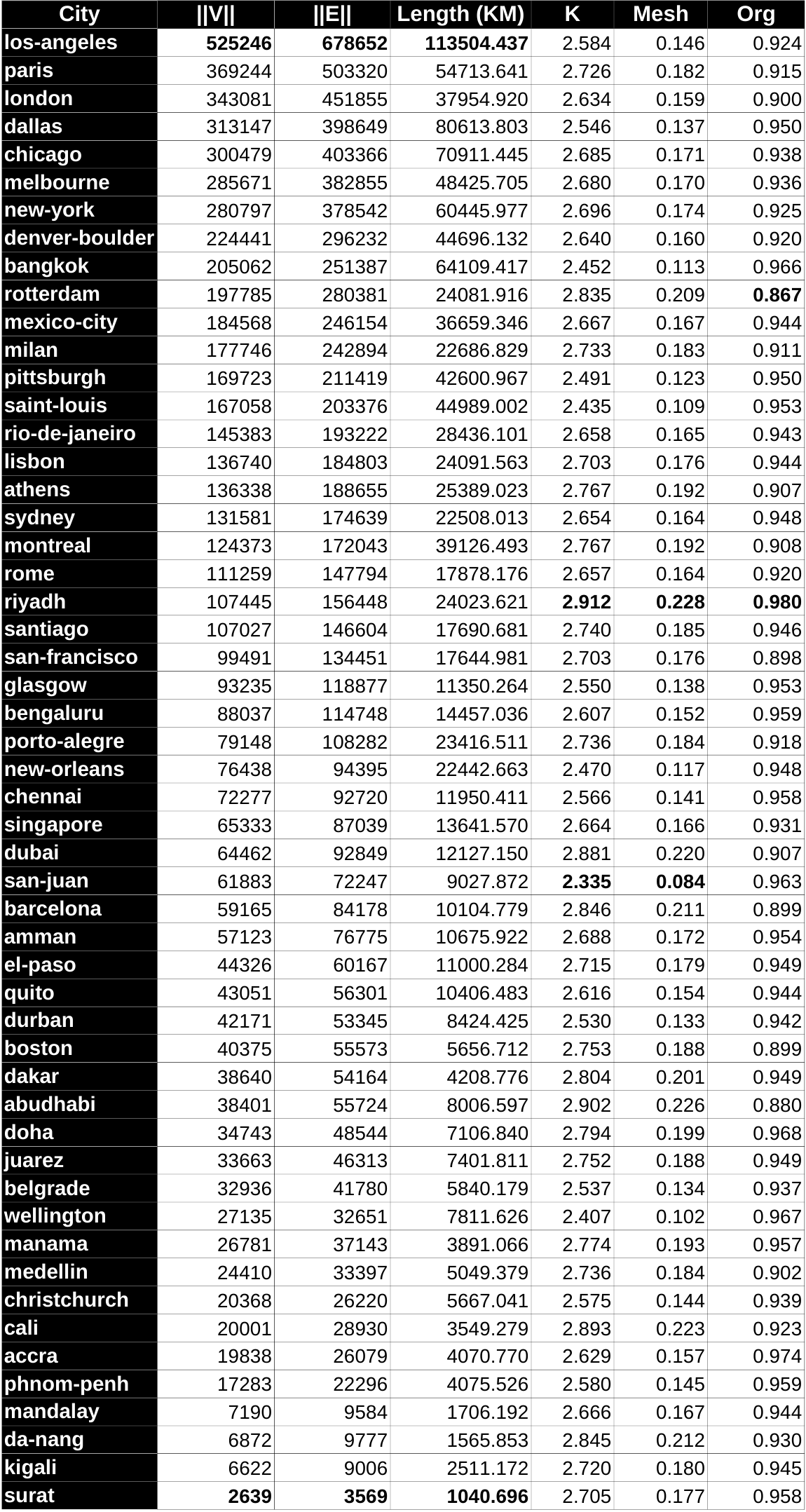}
\end{table}

\subsection{Service location data}
Foursquare is a location-based social media network that allows users to share their location with friends through what is known as {\em check-in}. Data from such check-ins contain information about the user, the venue she is reporting on, the specific location of that venue (coordinates) and a timestamp. 

In this work we exploit data that were collected by Yang {\em et al.} \cite{yang2015nationtelescope,yang2016participatory} from April 2012 to September 2013. This rich dataset contains 33,278,683 check-ins posted by 266,909 users on 3,680,126 unique venues, that are geographically spread in 415 cities across 77 countries. For our purposes, we only pay attention to the type of service the venue belongs to according to the Foursquare category hierarchy\footnote{https://developer.foursquare.com/categorytree}; and its GPS position --disregarding any other information. We have deliberately promoted the ``Medical Centers'' sub-category into a parent category to be able to assess how such a critical service is affected by failures in the city. At the end we obtained a list of 10 different categories of different criticality, including: {\em Medical Center, Travel \& Transport, Food, College \& University, Residence, Arts \& Entertainment, Shops \& service, Nightlife Spot, Professional \& Other Places, and Outdoors \& Recreation.}

To integrate the network structure with services locations, we assign each location to its closest node. This is tantamount to mapping service locations to their nearest intersection, allowing an efficient computation of the amount of reachable services (see Section 3). Services that are farther than 2 km for all intersections in a city are omitted from the study. 

\section{Methods}
Network percolation theory has already been exploited in the urban context for other purposes than the ones in this work \cite{arcaute2015hierarchical,li2015percolation,Jiaqiu2015Resilience}. With the road networks for dozens of cities at hand, we can now proceed with the percolation dynamics in two different ways. Both of them share the idea of {\em progressive structural deterioration} \cite{albert2000error,callaway2000network,cohen00}, understood either as error or failure (removal of randomly chosen edges); or attack (removal of important edges, where ``importance'' can be quantified by some descriptor, such as high betweenness of edges, high centrality of nodes, etc.) Note that in this work we focus on bond percolation (edges are removed) as opposed to site percolation (nodes are removed).

Here we apply both \texttt{Error} and \texttt{Attack} procedures on the available networks. For the \texttt{Error} process, at each step an edge chosen uniformly at random is removed, until the size of the GCC (giant connected component) is equal to the size of the SLCC (second larger connected component). Connected components (and their sizes) are checked periodically (each time 1\% of edges have been deleted). Because of the stochastic nature of this process, it is repeated 50 times so as to report the average expected deterioration of the network. In the \texttt{Attack} modality, the edge with highest betweenness centrality is removed at each step. As before, the process continues until the size of the GCC is equal to the size of the SLCC.

Following these progressive schemes, as the fraction $p$ of removed edges increases the size of the GCC (captured in number of nodes) decreases, until the network undergoes a second-order phase transition at the critical point $p_{c}$ (percolation threshold). Such transition can be spotted precisely --it occurs when the size of SLCC becomes maximal \cite{bollobas1985random}. 

For each network, at $p_{c}$ we compute the proportion of available services (gathered under ten general categories) in the remaining GCC as a proxy of the capacity of a city to deliver those services in an \texttt{Error/Attack} scenario.

\section{Results}
\begin{figure*}[htp]
 \centering
 \includegraphics[width=0.95\linewidth]{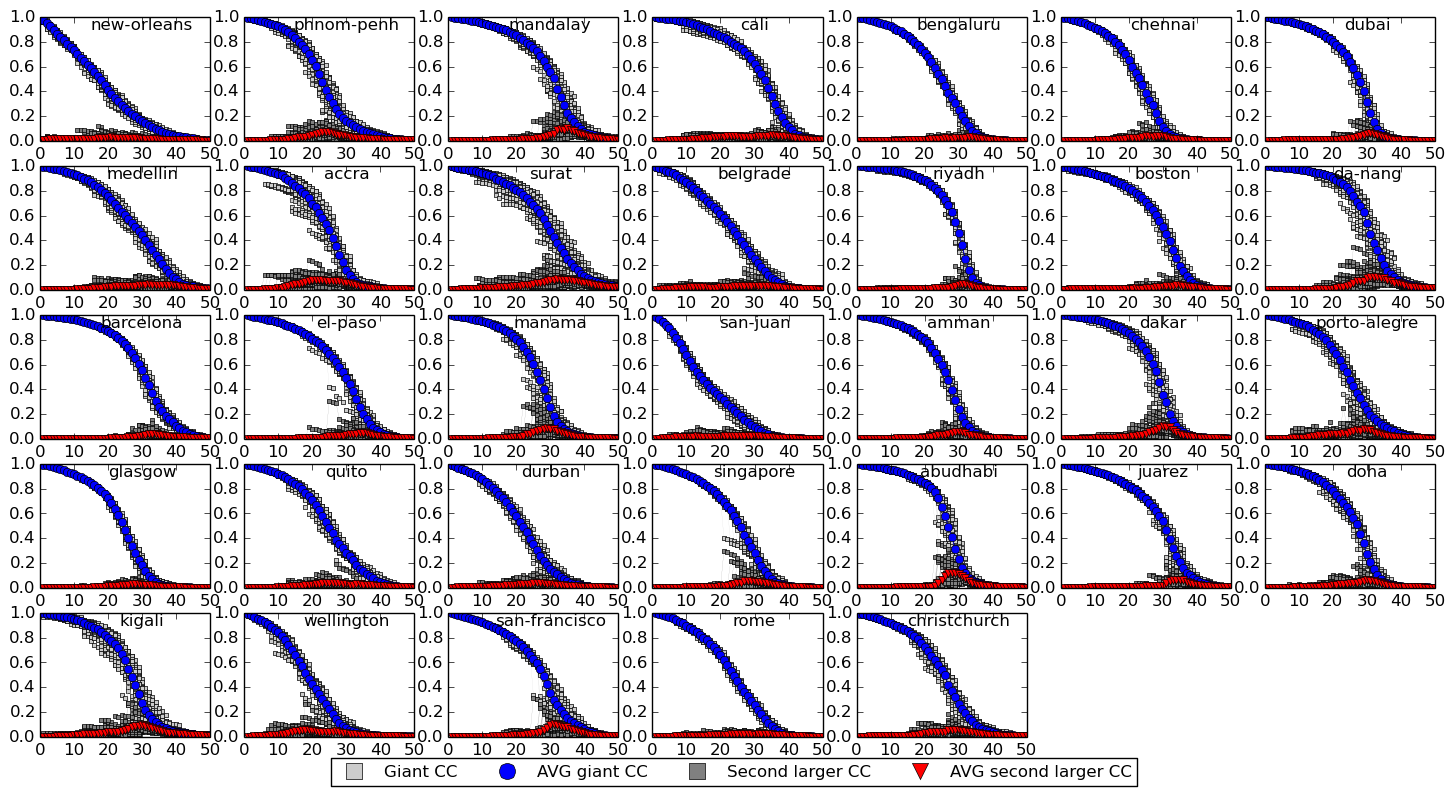}
 \caption{Random failure behavior. Evolution of the size of the GCC (y-axis) as a function of the percentage of removed edges (x-axis). Results are reported for 50 randomizations. Notably, cities react differently to edge failure due to their different $p_{c}$ --see Figure \ref{fig:percolation_threshold_comp} (left).}
 \label{fig:failure}
\end{figure*}

We first take a look at the cities' response to \texttt{Error}. Figure \ref{fig:failure} shows some examples of how random edge removal affects the size of the GCC. Typically, the size of the GCC is relatively stable, which indicates that the road structure is not suffering a large damage regarding overall reachability. At a given point ($p_{c}$), a transition takes place (often quite abruptly, e.g. Dakar or Abu-Dhabi), after which the network is shattered into many disconnected components. Note that \texttt{Error} results are only reported for a subset of cities to ease the comparison across different experiments.

\begin{figure*}[htp]
 \centering
 \subfigure[Random failure\label{fig:random_attack_bar}]{\includegraphics[width=1.\columnwidth]{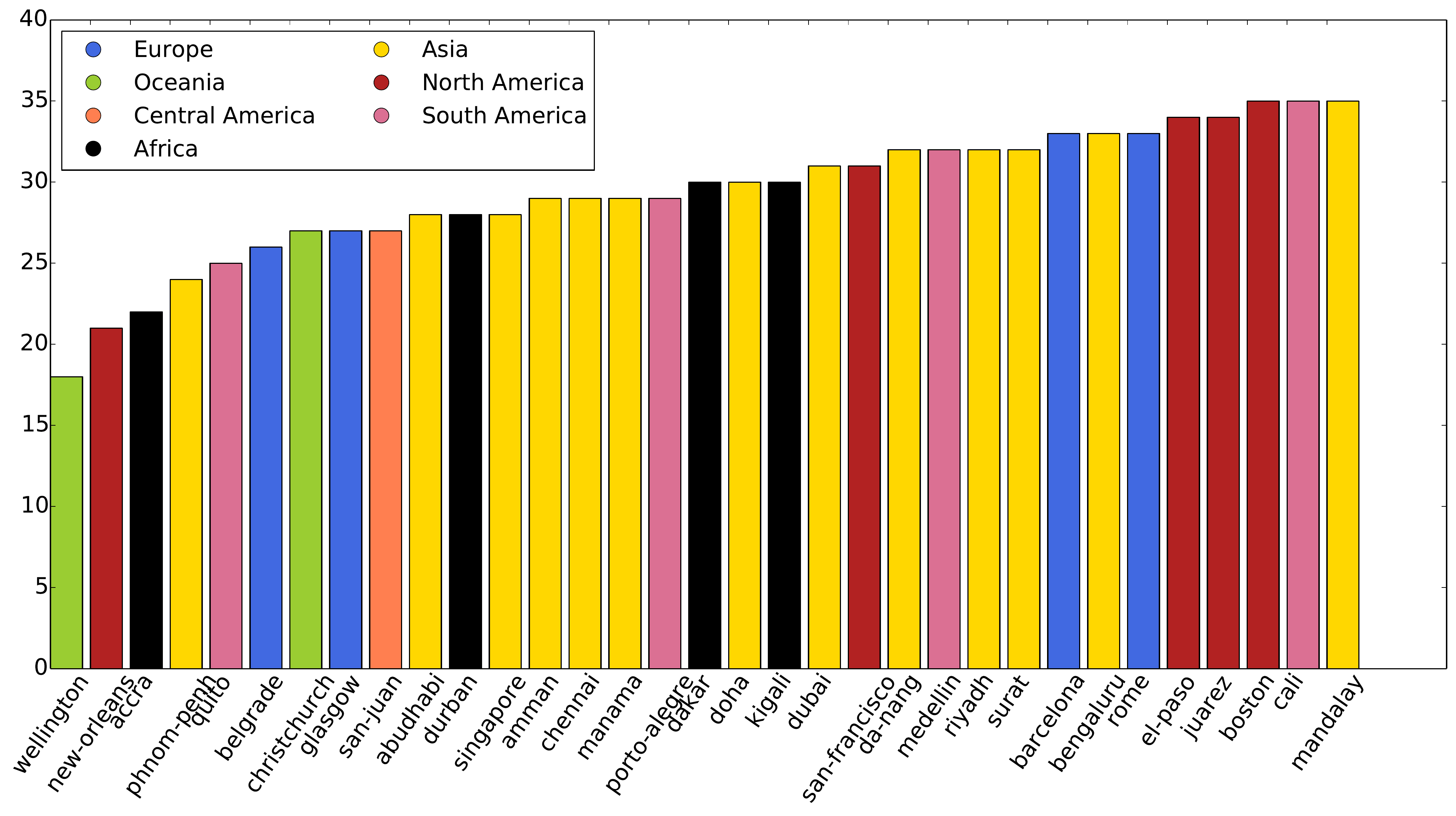}}
 \qquad
 \subfigure[Targeted attack\label{fig:intentional_attack_bar}]{\includegraphics[width=1.\columnwidth]{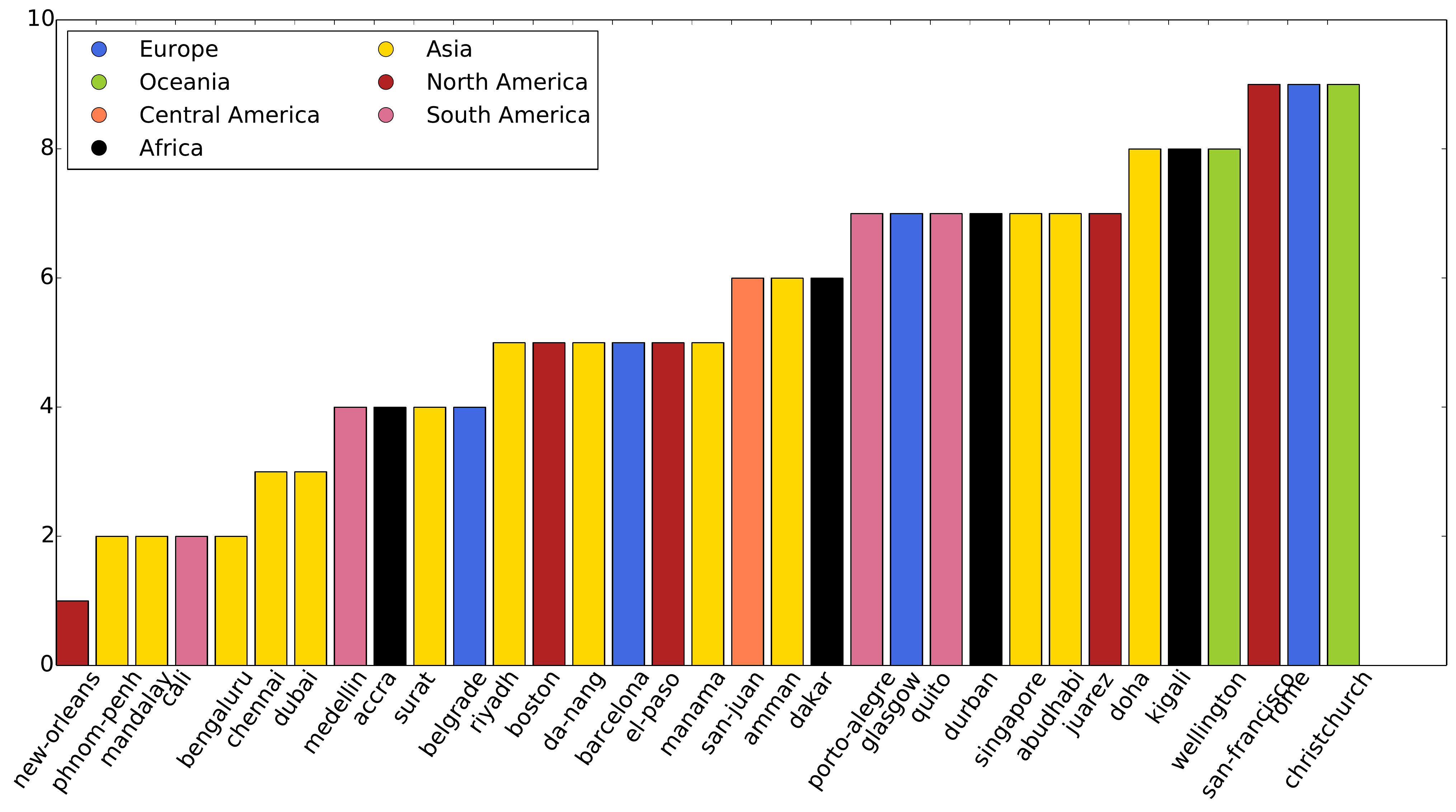}}
 \caption{Percolation point $p_{c}$ of different cities. The percolation point is computed as fraction of removed edges from the network when the size of the second larger connected component maximizes. Figure (a) reports the results for the random failure in which percolation thresholds are between 17\% (Wellington) and 35\% (Boston, Cali, and Mandalay). Figure (b) reports the results for the targeted attack. As expected, the percolation happens now much earlier, after removing only 1\% of edges in New Orleans, or 2\% in Phnom-Penh and Mandalay), to 9\% in San Francisco, Rome and Christchurch.}
 \label{fig:percolation_threshold_comp}
\end{figure*}

Along this process, $p_{c}$ can be seen as an indicator of the robustness of the system to failure. A low critical threshold can be interpreted as a limited capacity of the road infrastructure to support and recover from random disruption --accidents or congestion happening at any location, or system-wide disturbance. In Figure \ref{fig:percolation_threshold_comp} (left), cities are sorted according to their $p_{c}$ against random error (averaged over 50 realizations). We see there that there is a remarkably wide range of cities, with the weakest road network corresponding to Wellington ($p_{c}=0.17$), and the more robust ones to Mandalay, Cali and Boston ($p_{c}=0.35$). Such variability suggests deep structural differences beyond the obvious features reported in Table 1 (compare for instance the two mentioned cities). We also see that most North American have made it to the top cities in terms of robustness to random error. 

\begin{figure*}[htp]
 \centering
 \includegraphics[width=0.95\linewidth]{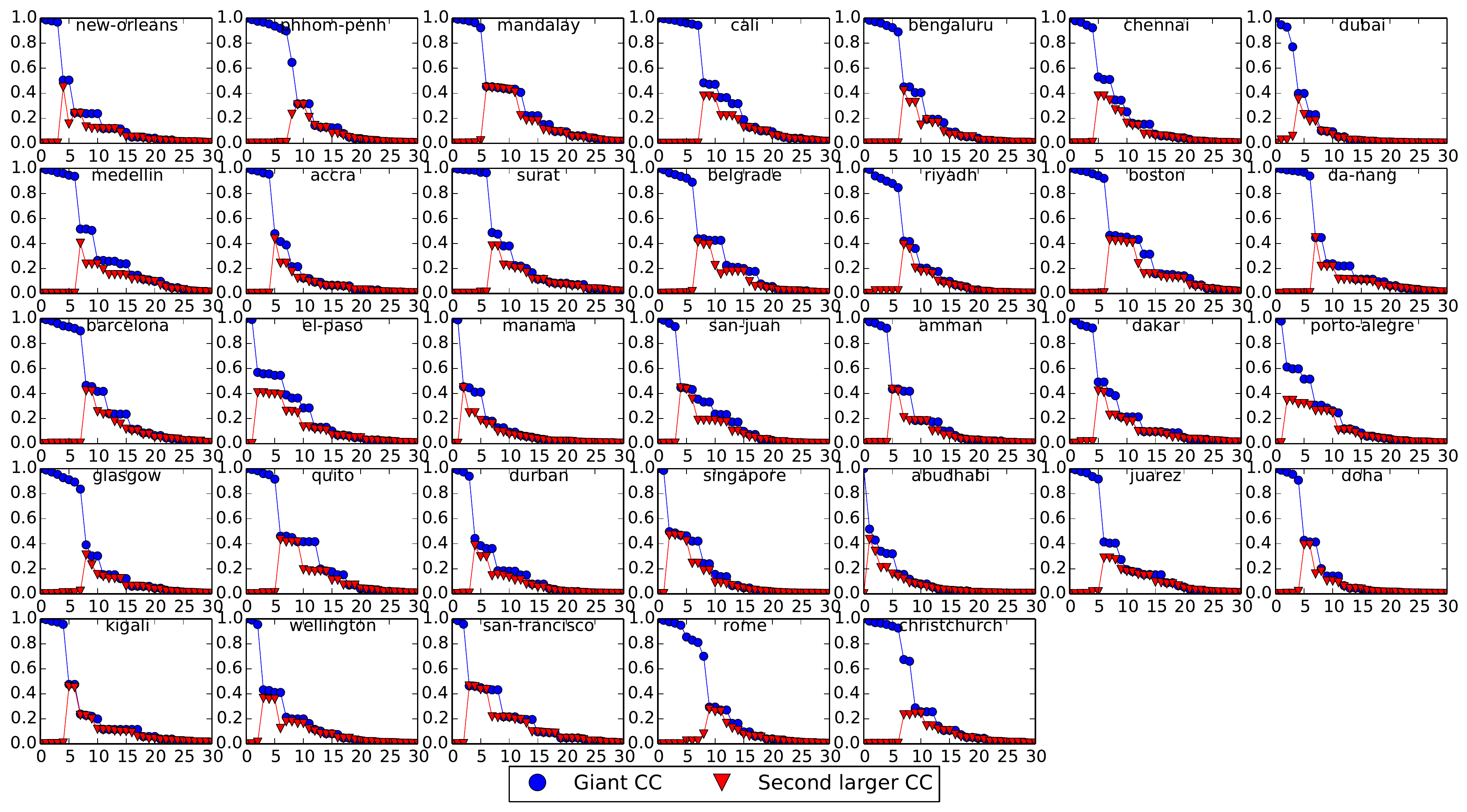}
 \caption{Intentional attack behavior: Evolution of the size of the GCC as a function of the percentage of removed edges. Again, cities react differently to edge failure --see Figure \ref{fig:percolation_threshold_comp} (right).}
 \label{fig:attack}
\end{figure*}

Figure \ref{fig:attack} shows as well the progressive deterioration of the GCC as a function of the fraction of removed edges, but in this case links are targeted with respect to their betweenness centrality. The differences with respect to Figure \ref{fig:failure} are obvious --transitions take place much earlier and in a much more abrupt manner. In this case, the values of $p_{c}$ are below 10\% for {\em any} city under study, evidencing the large dependence (and, in turn, exposure and vulnerability) of the road network on a few central segments. Figure \ref{fig:percolation_threshold_comp} (right) shows cities sorted by their percolation threshold under the targeted \texttt{Attack} scheme.

Interestingly, we see that the ranking of cities has drastically changed in \texttt{Attack}. In fact, cities from Oceania such as Wellington and Christchurch are turned to be among the most robust cities in these scheme while they were at the bottom of the ranking in the \texttt{Error} scheme. We also observe that North American cities are no longer grouped toward the head of the ranking as they were in the \texttt{Error} case. This suggest that cities, regardless of their location, behave differently the two failure schemes considered in our study. We further computed the overall Pearson correlation between the vectors of $p_{c}$ scores and we found it as low at $0.23$, indicating that there is a very low correlation between the percolation thresholds observed in \texttt{Error} and \texttt{Attack} schemes. In other words, a city with a high robustness to \texttt{Error} can be very vulnerable for \texttt{Attack} and vice versa.

\begin{figure*}[thp]
 \centering
 \subfigure[\label{fig:service_availability_bar}]{\includegraphics[width=.37\linewidth]{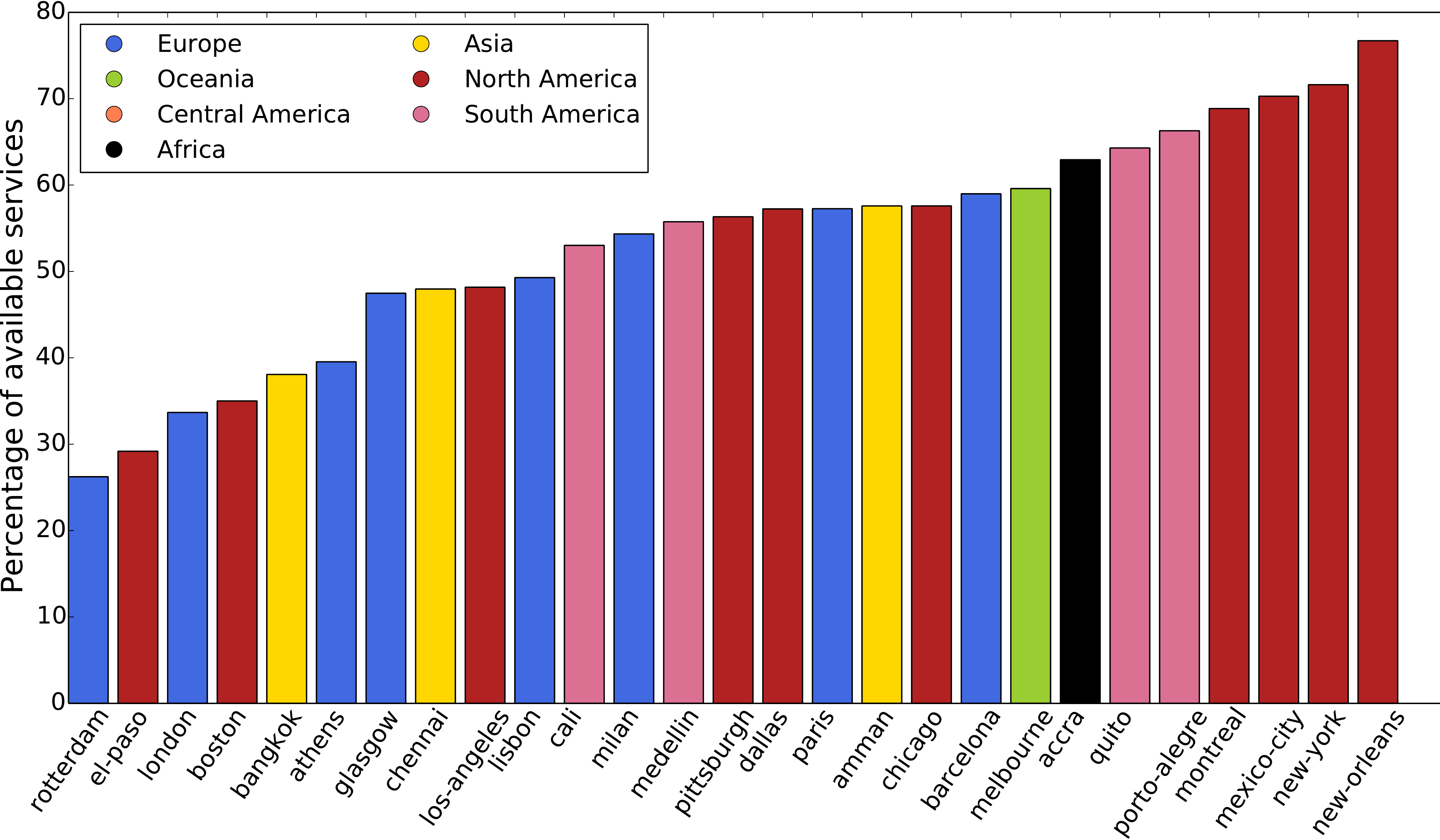}}
 \subfigure[\label{fig:service_availability_radar}]{\includegraphics[width=0.3\linewidth]{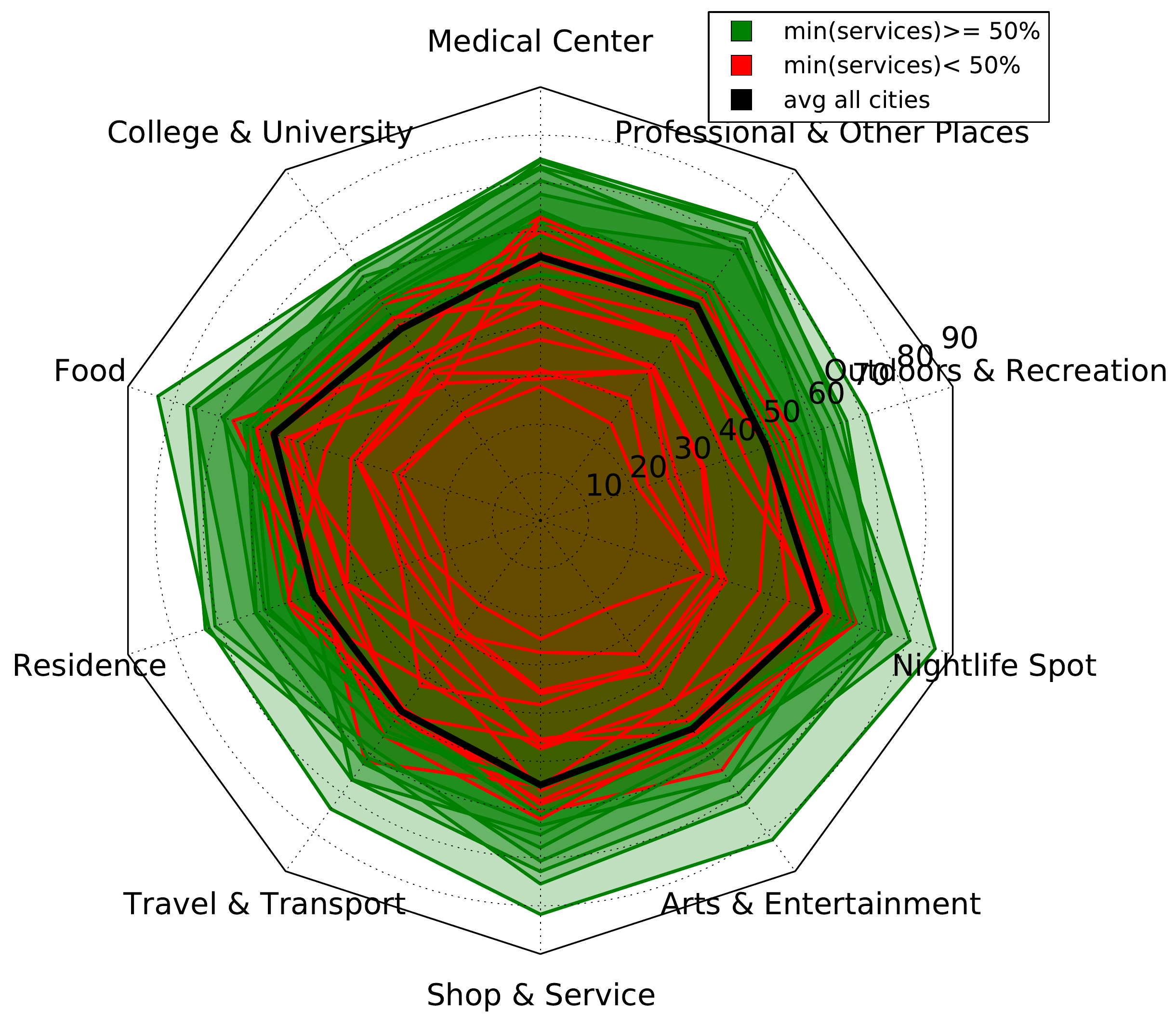}}
 \subfigure[\label{fig:service_availability_radar_eg}]{\includegraphics[width=0.3\linewidth]{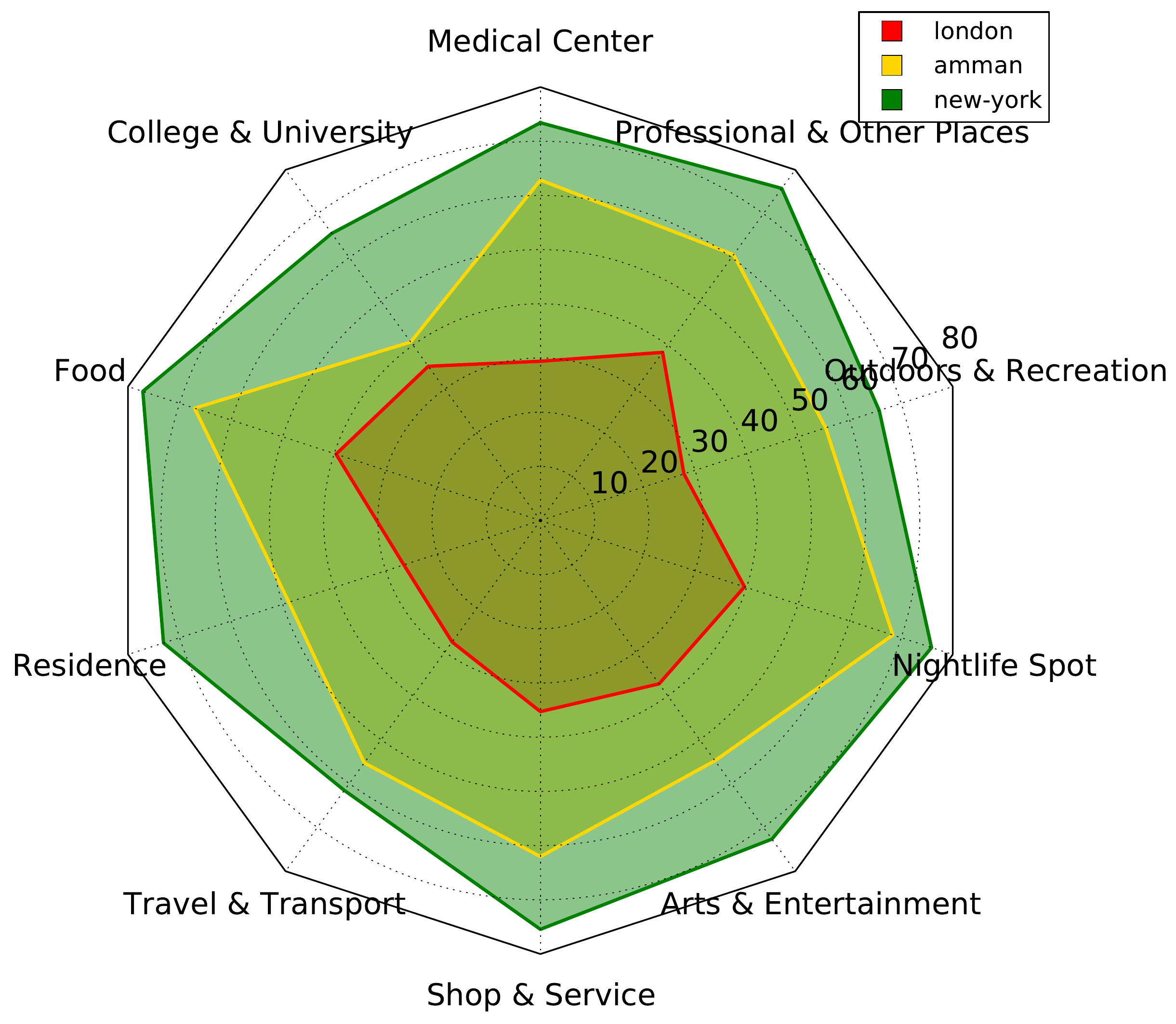}}
 \caption{Service availability in the giant connected component at percolation threshold $p_{c}$. Panel (a) reports the proportion of available services in different cities sorted from the most affected to the less affected. Panel (b): in green, cities that keep at least 50\% of their services in the GCC for all categories. In red, cities that lose more than 50\% services for at least one category. Panel (C): focus on three different cities with three different behaviors: London in red, Amman in yellow and New York in green}
 \label{fig:service_availability}
\end{figure*}

The third focus of our work is related to the question of service distribution. The effects of the failure of a part of the road network are not important {\em per se}, but in terms of the possibility to reach --or not-- any other part of the city (and the services it delivers). Thus, besides questioning how robust the infrastructure is (i.e. how early it breaks), we wonder now about the consequences of surpassing $p_{c}$ in terms of health, food or residential resources. Figure \ref{fig:service_availability_radar} summarizes the situation for the 27 cities for which we could find Foursquare data. In panel (a) of the same figure, we compute for each city the average proportion of available services (across the ten categories) reachable by the GCC observed at percolation threshold ($p_c$.) Here again, we found that North and South American cities such as New Orleans, New York, Motreal, Quito, and Porto-Alegre are in the top of the ranking. In the city of New Orleans for instance, only 23\% of the services became unaccessible (availability around 77\%.) The bottom of the ranking is fairly dominated by European cities with Rotterdam and London in the bottom 3 cities. Indeed, a city like Rotterdam could preserve access to only 26\% of all its services. Also,  there seem to be a negative correlation between the percolation threshold of a city and the fraction of services it retains. That is, the higher the percolation threshold the lower the available services. 

In panel (b) of Figure~\ref{fig:service_availability} we report the the availability of service per city per category. We represent in red the cities that lost more than 50\% of services in at least on category whereas we use green to report cities that preserve at least 50\% of their services in all categories. We also plot in black the average availability across cities for each category. As one would expect, cities are showing different patterns regarding different categories. However, a close analysis reveals that on average category availabilities are somewhat equilibrated around 55\% with the following breakdown: Nightlife Spot ($60.96\%$),  Food ($58.19\%$),  Professional \& Other Places ($55.18\%$), Shop \& Service ($54.89\%$), Medical Center($54.76\%$), Arts \& Entertainment ($53.50\%$), Residence ($49.51\%$), Outdoors \& Recreation ($49.36\%$), College~ \&~ University~ ($49.19\%$),~  Travel~ \&~ Transport ($49.00\%$). This reveals that the most affected category of services across cities is Travel and Transport that includes places such as bus and metro stations, airports, rail stations, etc. In the third and last panel (c) we report examples of cities with completely different reactions to failures in terms of service providing. London (UK) as a city that severely suffers the failure by losing access to almost 70\% of its services with a particular emphasis on medical centers and transport; Amman (Jordan) as a mid affected city losing access to 55\% of its services; and New York (US) -- as a good example -- looses only 30\% of its services with a high preservation of its medical centers.

\section{Discussion and Perspectives}
In this work, we have focused on various aspects of road network robustness. To do so, we have exploited large, publicly available datasets, and the rich tradition in percolation theory. Our results indicate that phase transitions occur at different levels of deterioration (early {\em vs.} late), but also in qualitatively different manners (abrupt {\em vs.} smooth). While these differences may result from different design principles \cite{Jiaqiu2015Resilience}, we suggest that these heterogeneous ``percolation profiles'' can be interpreted as a fingerprint of the city, in the sense that these profiles are unique to each of them --while at the same time incomplete (a city is much more than its road structure). 

We complement the structural degradation approach with a reachability perspective, coupling the infrastructural level with the services it connects. Indeed, a fractured road network is not critical {\em per se}, but in relation to the broken flow of people and resources. The results show again heterogeneous profiles regarding service allocation. These insights may be helpful in assessing strategies for a more balanced distribution of such services that guarantee reachability a balanced coverage even in a stressed infrastructure.

Despite the simplifications that our method entails, a systematic study of these ``robustness fingerprints'' may lead to a classification of cities with regard to their fragility and service distribution imbalances --along the lines of Louf \& Barthelemy \cite{louf2014typology}, but taking a urban resilience angle. Beyond its academic interest, such classification could lead to the design of common preventive strategies for cities whose structural features resemble each other (at least from a percolation and service allocation perspective). Furthermore, it could be enriched with other critical urban networks (e.g. water supply or power grid) expanding the scope of the study to multilayer percolation \cite{buldyrev2010catastrophic,gao2012networks}.

All of the previous falls under an overarching vision, which is related to the existing gaps between policy actors and scientific activity: the pipeline between both extremes is often too narrow --if not inexistent. There seems to be a general agreement that resilience is a desirable feature in urban settlements --as the world population increasingly flows into cities--, but the concept becomes fuzzy when it comes to the definition of objectives and strategies. In an attempt to alleviate this, we have taken a specific angle of resilience (that of 100RC) and mapped some of their targets onto quantifiable, concrete measurements. There is a large body of work waiting ahead before we can come up with an objective set of methods to measure urban resilience, taking into account the wide range of temporal and spatial scales a city encloses.

\section*{Acknowledgments}
We would like to thank Heather Leson for her discussions and comments that greatly improved the manuscript.

\bibliographystyle{abbrv}
\bibliography{cityRobustness}

\balancecolumns
\end{document}